\documentclass[aip,amsmath,amssymb,reprint]{revtex4-1}
\usepackage[table]{xcolor}
\usepackage{graphicx}% Include figure files
\usepackage{dcolumn}% Align table columns on decimal point
\usepackage{bm}% bold math
\usepackage[utf8]{inputenc}
\usepackage[T1]{fontenc}
\usepackage{mathptmx}
\usepackage{etoolbox}

\usepackage{tabularx,booktabs}

\makeatletter
\def\@email#1#2{%
 \endgroup
 \patchcmd{\titleblock@produce}
  {\frontmatter@RRAPformat}
  {\frontmatter@RRAPformat{\produce@RRAP{*#1\href{mailto:#2}{#2}}}\frontmatter@RRAPformat}
  {}{}
}%
\makeatother

\usepackage[capitalise]{cleveref}       % more clever refs
\usepackage{chemformula}    % chemical formulas
\usepackage[separate-uncertainty=true,exponent-product=\cdot]{siunitx}        % nicely typeset units
\DeclareSIUnit\angstrom{\text{\AA}} % angstrom is not a SI unit :clown_face:
\usepackage{graphicx}

\usepackage{mathtools}
\usepackage{todonotes}
\usepackage{xspace}
\usepackage{bbold} % to make a bold-faced 1
\usepackage{dsfont}

\usepackage[shortcuts]{glossaries}
\glsdisablehyper
\setacronymstyle{long-short}
\newacronym{gk}{GK}{Green-Kubo}
\newacronym{mic}{MIC}{minimum image convention}
\newacronym{dft}{DFT}{density-functional theory}
\newacronym{he}{HE}{Helfand-Einstein}
\newacronym{aigk}{aiGK}{\emph{ab initio} Green-Kubo}
\newacronym{md}{MD}{molecular dynamics}
\newacronym{aimd}{aiMD}{\emph{ab initio} molecular dynamics}
\newacronym{pes}{PES}{potential-energy surface}
\newacronym{ad}{AD}{automatic differentiation}
\newacronym{bte}{BTE}{Boltzmann transport equation}
\newacronym{hpc}{HPC}{high-performance computing}
\newacronym{hfacf}{HFACF}{heat flux autocorrelation function}
\newacronym[plural=MPNNs,
	firstplural=message-passing neural networks]
	{mpnn}{MPNN}{message-passing neural network}
\newacronym[plural=GLPs,
	firstplural=graph-based machine-learning potentials]
	{glp}{GLP}{graph-based machine-learning potential}
\newacronym[plural=MLPs,
	firstplural=machine-learning potentials]
	{mlp}{MLP}{machine-learning potential}
\newacronym{ffnn}{FFNN}{feed-forward neural network}
\newacronym{lammps}{LAMMPS}{Large-scale Atomic/Molecular Massively Parallel Simulator}
\newacronym{ase}{ASE}{Atomic Simulation Environment}
\newacronym{mae}{MAE}{mean absolute error}
\newacronym{mape}{MAPE}{mean absolute percentage error}
\newacronym{mp}{MP}{message-passing}
\newacronym{fcp}{FCP}{force constant potential}
\newacronym{ff}{FF}{forcefield}
\newacronym{eos}{EOS}{equation of state}

\newcommand{\mlp}{\gls{mlp}\xspace}
\newcommand{\gk}{\gls{gk}\xspace}
\newcommand{\mlps}{\glspl{mlp}\xspace}
\newcommand{\mpnn}{\gls{mpnn}\xspace}
\newcommand{\mpnns}{\glspl{mpnn}\xspace}

\newcommand{\ad}{\gls{ad}\xspace}
\newcommand{\glp}{\gls{glp}\xspace}
\newcommand{\glps}{\glspl{glp}\xspace}
\newcommand{\mic}{\gls{mic}\xspace}
\newcommand{\md}{\gls{md}\xspace}

\newcommand{\eos}{\gls{eos}\xspace}
\newcommand{\dft}{\gls{dft}\xspace}

\newcommand{\sok}{\textsc{so3krates}\xspace}
\newcommand{\nequip}{\textsc{nequip}\xspace}
\newcommand{\cglp}{\texttt{glp}\xspace}

\newcommand{\cmlff}{\texttt{mlff}\xspace}
\newcommand{\jax}{\texttt{jax}\xspace}

\newcommand{\SM}{Supp. Mat.\xspace}

\newcommand{\scare}[1]{\lq#1\rq} 
\newcommand{\etal}{et~al.}

\newcommand{\mel}{\,...\,\xspace}
\newcommand{\symset}[1]{\mathrm{#1}}  % symbol of a set

\newcommand{\tk}{\boldsymbol{\kappa}}
\newcommand{\J}{\boldsymbol{J}}
\newcommand{\Jfull}{\boldsymbol{J}_\text{full}}

\newcommand{\Jconv}{\J_{\text{convective}}}

\newcommand{\td}{t_0}

\newcommand{\dur}[3]{\frac{\partial U_{#1}}{\partial \R_{#2#3}}}

\newcommand{\nbh}[1]{\symset{N}(#1)}

\newcommand{\interactions}{M}
\newcommand{\cutoff}{r_{\text{c}}}
\newcommand{\effcutoff}{\cutoff^{\text{eff}}}

\newcommand{\graph}{\symset{G}}
\newcommand{\vertices}{\symset{V}}
\newcommand{\edges}{\symset{E}}

\newcommand{\R}{\boldsymbol{r}}

\newcommand{\Rm}{\boldsymbol{r}^{\text{MIC}}}

\newcommand{\V}{\boldsymbol{v}}

\newcommand{\F}{\boldsymbol{F}}
\newcommand{\Bary}{\boldsymbol{B}}

\newcommand{\defas}{\coloneqq}  % needs \usepackage{mathtools}
\newcommand{\defdas}{\eqqcolon}

\newcommand{\curlyset}[2]{\{\, #1\, :\, #2 \, \}}

\newcommand{\wholes}{\mathbb{Z}}
\newcommand{\reals}{\mathbb{R}}

\newcommand{\siy}{\mathbb{1}}

\newcommand{\magnitude}[1]{\bigl|#1\bigr|}

\newcommand{\stress}{\boldsymbol{\sigma}}

\newcommand{\strain}{\boldsymbol{\epsilon}}

\newcommand{\Z}{\symset{Z}}
\newcommand{\Us}{\symset{U}}

\newcommand{\basis}{\boldsymbol{b}}
\newcommand{\offset}{\boldsymbol{n}}

\newcommand{\Rsc}{\symset{R_{\text{sc}}}}

\newcommand{\Rall}{\symset{R_{\text{all}}}}
\newcommand{\Basis}{\symset{B}}

\newcommand{\Runf}{\symset{R_{\text{unf}}}}

\newcommand{\gstd}{\graph^{\text{std}}}
\newcommand{\gunf}{\graph^{\text{unf}}}
\newcommand{\estd}{\edges^{\text{std}}}

\newcommand{\nbht}[1]{\symset{N}^\interactions\!\!(#1)}
\newcommand{\Jpot}{\boldsymbol{J}_\text{pot}}
\renewcommand{\Jconv}{\boldsymbol{J}_\text{conv}}
\newcommand{\Rng}{\boldsymbol{r}^\text{aux}}

\newcommand{\vecx}{\boldsymbol{x}}

\definecolor{yellow2}{HTML}{FADF63}

\newcommand{\maxf}[1]{{\cellcolor[gray]{0.8}} #1}

\begin{document}

\preprint{AIP/123-QED}

\title{Stress and heat flux via automatic differentiation}

\begin{abstract}
Machine-learning potentials provide computationally efficient and accurate approximations of the Born-Oppenheimer potential energy surface. This potential determines many materials properties and simulation techniques usually require its gradients, in particular forces and stress for molecular dynamics, and heat flux for thermal transport properties.
Recently developed potentials feature high body order and can include equivariant semi-local interactions through message-passing mechanisms.
Due to their complex functional forms, they rely on automatic differentiation (AD), overcoming the need for manual implementations or finite-difference schemes to evaluate gradients.
This study demonstrates a unified AD approach to obtain forces, stress, and heat flux for such potentials, and provides a model-independent implementation.
The method is tested on the Lennard-Jones potential, and then applied to predict cohesive properties and thermal conductivity of tin selenide using an equivariant message-passing neural network potential.
\end{abstract}

\author{Marcel F. Langer}
\email{mail@marcel.science}
\affiliation{Machine Learning Group, Technische Universit{\"a}t Berlin, 10587 Berlin, Germany}
\affiliation{BIFOLD -- Berlin Institute for the Foundations of Learning and Data, Berlin, Germany}
\affiliation{The NOMAD Laboratory at the Fritz Haber Institute of the Max Planck Society and Humboldt University, Berlin, Germany}

\author{J. Thorben Frank}
\affiliation{Machine Learning Group, Technische Universit{\"a}t Berlin, 10587 Berlin, Germany}
\affiliation{BIFOLD -- Berlin Institute for the Foundations of Learning and Data, Berlin, Germany}

\author{Florian Knoop}
\affiliation{Theoretical Physics Division, Department of Physics, Chemistry and Biology (IFM), Linköping University, SE-581 83 Linköping, Sweden}

\date{\today}%

\maketitle

\section{Introduction}

\Gls{md} simulations enable computational prediction of thermodynamic quantities for a wide range of quantum systems, and constitute a cornerstone of modern computational science~\cite{tuckerman2010}.
In \md, systems are simulated by propagating Newton's equations of motion -- potentially modified to model statistical ensembles -- numerically in time, based on the forces acting on each atom due to their movement on the Born-Oppenheimer \gls{pes}.
Therefore, the quality of the underlying \gls{pes} is important for the predictive ability of this method.
First-principles electronic structure methods such as \gls{dft} can be used to perform high-accuracy \md simulations~\cite{cp1985p}, provided the exchange-correlation approximation is reliable~\cite{thxy2022p}.
Such approaches are, however, restricted by high computational cost, severely limiting accessible size and time scales.
Computationally efficient approximations to the underlying \gls{pes} are therefore required for the atomistic simulation of larger systems:
\Glspl{ff} are built on analytical functional forms that are often based on physical bonding principles, parametrized to reproduce quantities of interest for a given material~\cite{g2011p}. They are computationally cheap, but parametrizations for novel materials are not always available, and their flexibility is limited by their fixed functional form.
\Glspl{mlp}~\cite{lgs2004q,lhsr2006q,bp2007q,bpkc2010q,lgs2004q,bp2007q,csmt2018q,csmt2019q,pt2021q}, where a potential is inferred based on a small set of reference calculations, aim to retain near first-principles accuracy while remaining linear scaling with the number of atoms.
While \mlps are limited, in principle, to modeling the physical mechanisms present in the training data, they have nevertheless emerged as an important tool for \md~\cite{ntmc2020q,ukkm2020q,kvmt2021q,uctm2021q,ustm2022a}, often combined with active learning schemes~\cite{capd2004q,jmka2020q,vkjk2021q,osoc2022a,xvjk2022a}.
Modern \mlps can include semi-local interactions through \gls{mp} mechanisms~\cite{gsvd2017q}, internal equivariant representations~\cite{tskr2018q}, and body-order expansions~\cite{d2019q}, which enable the efficient construction of flexible many-body interactions.
In such complex architectures, the manual implementation of derivatives is often unfeasible. Finite-difference approaches require tuning of additional parameters, as well as repeated energy evaluations.
\Gls{ad}~\cite{griewank2008,bprs2017m} presents an intriguing alternative:
If the computation of the potential energy $U$ is implemented in a suitable framework, derivatives such as the forces $\F$ or the stress $\stress$ can be computed with the same asymptotic computational cost as computing the energy.
This is accomplished by decomposing the underlying \scare{forward} computation into elementary operations with analytically tractable derivatives, computing local gradients for a given input, and then combining the resulting values using the chain rule. 

This work provides a systematic discussion of the use of \ad to compute forces, stress, and heat flux for \mlps.
While the calculation of forces with \ad is common practice~\cite{jax-md,sktm2018q,um2019q,shlg2023q}, stress and heat flux are not yet commonly available for many \mlps.
Both quantities have been the focus of much previous work due to the difficulty of defining and implementing them for many-body potentials and periodic boundary conditions~\cite{lb2006p,tno2008t,tpm2009t,vc1999t,c2006t,ggs2010t,fpdh2015t,mub2015t,crs2017t,bbw2019t,smko2019t,lksr2023a}.
Introducing an abstract definition of \mlps as functions of a graph representation of atomistic systems, unified formulations of stress and heat flux are given, which can be implemented generally for any such \glp.
An example implementation using \jax~\cite{jax} is provided in the \cglp package~\cite{glp}.
To validate the approach, different formulations of stress and heat flux are compared for the Lennard-Jones potential~\cite{l1924p}, where analytical derivatives are readily available for comparison, as well as a state-of-the art \mpnn, \sok~\cite{fum2022q}.
Having established the correctness of the proposed implementation, the ability of \sok to reproduce first-principles cohesive properties and thermal conductivity of tin selenide (\ch{SnSe}) is studied. 

\section{Automatic Differentiation}

\emph{Automatic differentiation} (\ad)  is a technique to obtain derivatives of functions implemented as computer programs.~\cite{griewank2008,bprs2017m} It is distinct from \emph{numerical differentiation}, where finite-difference schemes are employed, and \emph{symbolic differentiation}, where analytical derivatives are obtained manually or via computer algebra systems, and then implemented explicitly. Instead, \ad relies on the observation that complex computations can often be split up into elementary steps, for which derivatives are readily implemented. If one can track those derivatives during the computation of the forward, or \scare{primal}, function, the chain rule allows to obtain derivatives. 

For this work, two properties of \ad are particularly relevant: It allows the computation of derivatives with respect to quantities that are explicitly used in the forward computation, and it can do so at the same asymptotic computational cost as the forward function.
In particular, \ad can obtain two quantities efficiently: Given a differentiable function $u: \reals^N \rightarrow \reals^M$, the Jacobian of $u$ is defined as the $M \times N$ matrix $\partial u_i(\vecx)/\partial x_j$. 
\ad can then obtain Jacobian-vector and vector-Jacobian products, i.e., the multiplication and summation of factors over either the input or the output dimension.
This corresponds to propagating derivatives from the inputs \emph{forwards}, leading to \emph{forward-mode \ad}, or from the end result \emph{backwards}, leading to \emph{reverse-mode \ad}.
As many popular \ad frameworks are primarily implemented to work with neural networks, where scalar loss functions must be differentiated with respect to many parameters, reverse-mode \ad, also called \scare{backpropagation},~\cite{rhw1986m} is more generally available. More recent frameworks implement both approaches, for instance \jax~\cite{jax} which is used in the present work.
\ad can also be leveraged to compute contractions of higher-order derivative operators~\cite{smc2022q}.
% As an extreme case, if $M=1$, the gradient of $u$ can therefore be obtained efficiently. 

\section{Constructing Graph MLPs}

% Since \ad relies on the forward computation to calculate derivatives, it is sensitive to the exact implementation of a given potential energy function. Care must therefore be taken to construct the \mlp accordingly, ensuring that required derivatives are available.

This work considers periodic systems,\footnote{Non-periodic systems can be formally accommodated by setting $\Basis$ such that replicas lie outside the interaction cutoff radius. The \cglp framework supports non-periodic systems.} consisting of $N$ atoms with atomic numbers $Z_i$ placed in a simulation cell which is infinitely periodically tiled in space. We define
\begin{alignat}{2}
	\Rsc &\defas \curlyset{\R_i}{i = 1 ... N} & \text{positions in simulation cell}  \nonumber\\
	\Z &\defas \curlyset{Z_i}{i = 1 ... N} & \text{atomic numbers}  \nonumber\\
	\Basis &\defas \curlyset{\basis_a}{a = 1, 2, 3} & \text{basis or lattice vectors}  \nonumber\\
	\R_{i\offset} &\defas \R_i + \sum_a\nolimits n_a \basis_a & \text{(replica) position} \nonumber\\
	% \Rrep &= \curlyset{\R_{i\offset}}{\R_i \in \Rsc, \offset \in \wholes^3, \offset \neq \boldsymbol{0}} & \text{replica positions}  \nonumber\\
	\Rall &\defas \curlyset{\R_{i\offset}}{\R_i \in \Rsc, \offset \in \wholes^3} & \text{all (bulk) positions} \nonumber \\
	\R_{ij} &\defas \R_{j} - \R_{i} & \text{atom-pair vector} \nonumber \\
	\magnitude{\Rm_{ij}} &\defas \min_{\offset \in \mathbb{Z}^3} \magnitude{\R_j + \sum_a n_a \basis_a - \R_i} & \text{minimum image convention} \nonumber
\end{alignat}
In this setting, a \mlp{} is a function that maps the structure represented by its positions, lattice vectors, and atomic numbers, $(\Rsc, \Basis, \Z)$, to a set of atomic potential energies $\Us \defas \curlyset{U_i}{i=1\mel N}$, which yield the total potential energy $U = \sum_{i=1}\nolimits U_i$.
Since \ad relies on the forward computation to calculate derivatives, it is sensitive to the exact implementation of this mapping. Care must therefore be taken to construct the \mlp such that required derivatives are available.

This work considers \mlps that scale linearly with $N$. Therefore, the number of atoms contributing to a given $U_i$ must be bounded, which is achieved by introducing a cutoff radius $\cutoff$, restricting interactions to finite-sized atomic neighborhoods $\nbh{i} = \curlyset{\R_j}{\bigl|\R_{ij}\bigr| \leq \cutoff, \R_j \in \Rall}$.
To ensure translational invariance, \mlps{} do not rely on neighbor positions directly, but rather on atom-pair vectors centered on $i$, from which atom-pair vectors between neighboring atoms can be constructed, for instance to determine angles.

The resulting structure can be seen as a graph $\graph$. The vertices $\vertices$ of this graph are identified with atoms, labeled with their respective atomic numbers, and connected by edges $\edges$ that are labeled with atom-pair vectors if placed closer than $\cutoff$.
Starting from $\graph$, \mlps{} can be constructed in different ways: Local \mlps{} compute a suitable representation~\cite{lgr2022q}
% , ensuring rotational and permutational invariance, 
of each neighborhood, and predict $U_i$ from that representation using a learned function, such as a neural network or a kernel machine. Such models are conceptually simple, but cannot account for effects that extend beyond $\cutoff$. Recently, semi-local models such as \mpnns~\cite{gsvd2017q,sktm2017q,sstm2018q,um2019q,kgg2020q,bmsk2022q,ucsm2021q,bkoc2022q,bbkc2022a,blcd2022q} have been introduced to tackle this shortcoming without compromising asymptotic runtime. In such models, effective longer-range interactions are built up iteratively by allowing adjacent neighborhoods to interact repeatedly. We introduce the parameter $\interactions$, the \emph{interaction depth}, to quantify how many such iterations are included.
After $\interactions$ interactions, the energy at any given site can depend implicitly on positions within $\interactions$ hops on the graph, which we denote by $\nbht{i}$, leading to an \emph{effective} cutoff radius $\effcutoff = \interactions \, \cutoff$.
However, since interactions are confined to neighborhoods at each iteration, the asymptotic linear scaling is not impacted. Local \mlps are formally included as semi-local models with $\interactions{=}1$, allowing a unified treatment for both. We term this class of potentials, which act on sets of neighborhoods and use atom-pair vectors as input, \emph{graph-based machine-learning potentials (\glps)}.
By construction, this framework does not include global interactions.

We consider two strategies to construct \glps: The \scare{standard} way, which includes periodic boundary conditions via the edges in the graph, and an \scare{unfolded} formulation, where periodicity is explicitly included via replica positions.

In the standard architecture, vertices in $\graph$ are identified with atoms \emph{in the simulation cell}, using the \mic to include periodicity. Edges $\edges_{ij}$ exist between two atoms $i$ and $j$ in $\Rsc$ if they interact:
\begin{equation}
	\estd_{ij} = \curlyset{\Rm_{ij}}{\magnitude{\Rm_{ij}} \leq \cutoff} \, .
\end{equation}
We denote the the graph constructed in this manner as $\gstd$ and the set of edges $\estd$.

Alternatively, we can first determine the total set of positions $\Runf \subset \Rall$ that can interact with atoms in the simulation cell, creating an \scare{unfolded} system extracted from the bulk, consisting of $\Rsc$ and all replicas with up to $\effcutoff$ distance from the cell boundary. This construction can be performed efficiently, and adds only a number of positions that is proportional to the \emph{surface} of the simulation cell, therefore becoming increasingly negligible as $N$ increases at constant density~\cite{lksr2023a}.
We proceed by constructing a correspondingly modified graph $\gunf$, and compute potential energies for vertices corresponding to atoms in the simulation cell only. By construction, since the same atom-pair vectors appear in the graph, this approach then reproduces the potential energy of the standard method.
% TODO: note somewhere that this is because replicas are symmetrically equivalent and so all embeddings etc must be the same as the other case

\section{Derivatives}

Having constructed the forward function for a given \glp{}, we can compute derivatives with respect to its inputs using \ad. In this section, we discuss how forces, stress, and heat flux can be computed in this manner, and demonstrate the relationship between different formulations.

\subsection{Forces}

For \md, the most relevant quantities are the forces
\begin{equation}
	\F_i =-\dur{}{i}{}
\end{equation}
acting on the atoms in the simulation cell. Since $\Rsc$ are an explicit input, they can be computed directly with \ad: $U$ is a scalar, and this therefore is a trivial Jacobian-vector product, which can be computed with the same asymptotic cost as $U$.

An interesting situation arises if pairwise forces are desired. Strictly speaking, in a many-body \mlp{}, where interactions cannot be decomposed into pairwise contributions, such quantities are not well-defined and Newton's third law is replaced by conservation of momentum, which requires $\sum_{i=1}^N\nolimits \F_i = 0$. Nevertheless, pairwise forces with an antisymmetric structure can be defined by exploiting the construction of \glps{} in terms of atom-pair vectors. In the standard formulation, $U$ is a function of all edges,
\begin{equation}
	U = U(\curlyset{\R_{ij}}{ij \in \estd}) \, .
\end{equation}
Hence, by the chain rule,
\begin{align}
	\F_i &= \sum_{j \in \nbh{i}} \dur{}{i}{j} - \dur{}{j}{i} \\
	&\defdas \sum_{j \in \nbh{i}} \F_{ij} \, .
\end{align}
The pairwise forces such defined exhibit anti-symmetry, and therefore fulfil Newton's third law.
% Since $U$ depends on \emph{all} positions, changes in positions further apart than $\effcutoff$ from $\R_i$ can change $\F_i$.
For $\interactions{=}1$, the local case, this definition reduces to a more standard form~\cite{fpdh2015t}
\begin{equation}
	\F_{ij} = \dur{i}{i}{j} - \dur{j}{j}{i} \, .
\end{equation}
However, for general \glps{} with $\interactions{>}1$, this definition includes a sum over all $U_k$ that are influenced by a given edge
\begin{equation}
	\F_{ij} = \sum_{k \in \nbht{i}} \dur{k}{i}{j} - \dur{k}{j}{i} \, ,
\end{equation}
subverting expectations connecting local potential energies to pairwise forces.
We note that this seeming contradiction is a consequence of the combination of the peculiar construction of \glps and \ad: In principle, it is always possible to define extended neighborhoods up to $\effcutoff$, obtaining $U_i$ purely as a function of atom-pair vector originating from $i$. However, to construct derivatives with respect to these atom-pair vectors with \ad, these extended neighborhoods have to be constructed and included explicitly, therefore negating the computational efficiency gains of a \glp architecture.

\subsection{Stress}
\label{sec:stress}

The definition of the (potential) stress is~\cite{kcbs2015t}
\begin{equation}
	\stress = \frac{1}{V} \, \frac{\partial U}{\partial \strain}{\big |}_{\strain = 0} \, ,
	\label{eq:stress}
\end{equation}
with $U$ denoting the potential energy after a strain transformation with the the $3\times3$ symmetric tensor $\strain$
\begin{equation}
    \R \rightarrow (\siy + \strain) \cdot \R \, ,
\end{equation}
acting on $\Rall$.
While computing this derivative for arbitrary potentials and periodic systems has required \scare{much effort}~\cite{lb2006p} in the past~\cite{tpm2009t,at2011t}, it is straightforward with \ad.

The simplest approach, followed for instance by \texttt{schnetpack}~\cite{sktm2018a,schnetpack} and \texttt{nequip}~\cite{bmsk2022q,nequip}, is to inject the strain transformation explicitly into the construction of the \glp.
This can be done at different points: One can transform $\Rsc$ and $\Basis$ \emph{before} constructing $\graph$, directly transform atom-pair vectors, or transform all contributing positions $\Runf$.
Alternatively, as the inner derivative of inputs with respect to $\strain$ is simply the input, the derivative of $U$ with respect to inputs can be obtained with \ad, and the stress computed analytically from the results. This avoids modifying the forward computation of $U$ entirely.

These approaches yield, with $\otimes$ denoting an outer product,
\begin{align}
	V \stress
	&=\frac{\partial U(\Rsc,\Basis)}{\partial \strain} \label{eq:glp_stress_strain_direct}\\
    &=\frac{\partial U(\edges)}{\partial \strain} \label{eq:glp_stress_strain_edges}\\
    &=\frac{\partial U(\Runf)}{\partial \strain} \label{eq:glp_stress_strain_unf}\\
    &=\sum_{i \in \Rsc} \R_j \otimes \dur{}{i}{} + \sum_{\basis \in \Basis} \basis \otimes \frac{\partial U}{\partial \basis} \label{eq:glp_stress_sc_basis}\\
    &= \sum_{ij \in \edges} \R_{ij} \otimes \dur{}{i}{j} \label{eq:glp_stress_edges}\\
    &= \sum_{i \in \Runf} \R_{i} \otimes \dur{}{i}{} \label{eq:glp_stress_bulk} \, ,
\end{align}
recovering previous formulations given by Louwerse and Baerends~\cite{lb2006p} and Thompson~\cite{tpm2009t}. As seen in \cref{tab:stress.lj,tab:stress.snse}, all such forms of the stress are equivalent, provided the strain transformation is applied consistently to all used inputs.\footnote{For instance, \cref{eq:glp_stress_strain_direct} and \cref{eq:glp_stress_sc_basis} yield incorrect results if the \mic is implemented using fractional coordinates, rather than the definition using offsets.}
In all cases, as $U$ is differentiated with respect to its inputs, asymptotic cost remains linear.

A more complex situation arises if strain derivatives of atomic energies, i.e., atomic stresses
\begin{equation}
	\stress_i \defas \frac{1}{V} \frac{\partial U_i}{\partial \strain} \hspace{1cm} i \in \Rsc \label{eq:stress_atomic}
\end{equation}
are required.
Their calculation requires either one backward pass per $U_i$, or one forward pass for each entry in $\strain$. If only reverse-mode \ad is available, its evaluation therefore scales quadratically with $N$.  Linear scaling is retained with forward mode. 
For \glps with $\interactions{=}1$, linear scaling in reverse mode can be recovered by using \cref{eq:glp_stress_edges}: Every edge can be uniquely assigned to one $U_i$, and therefore the derivatives can be used to construct atomic stresses. For $\interactions{>}1$, this is not possible; similar to the observations of the previous section, atomic stresses take a semi-local form.

\subsection{Heat Flux}

Finally, we discuss the heat flux, which is required to compute thermal conductivities with the \gls{gk} method~\cite{g1952t,k1957t,kyn1957t}. It describes how energy flows between atoms, and has been the focus of a large body of previous work~\cite{c2006t,at2011t,tno2008t,fpdh2015t,smko2019t,bbw2019t,lksr2023a}.

The fundamental definition of the heat flux for \mlps was originally derived by Hardy~\cite{h1963t} for periodic quantum systems. It reads~\cite{lksr2023a}
\begin{align}
    \Jfull 
        &= \sum_{\substack{i \in \Rsc \\ j \in \Rall}} \left(\R_{ji} \left(\dur{i}{j}{} \cdot \V_j\right) \right) 
        + \sum_{i \in \Rsc} E_i \V_i \label{eq:hf_general}\\
        &\defdas \Jpot + \Jconv\, , 
\end{align}
where $\V_i$ denote velocities, $m_i$ masses, and $E_i = U_i + 1/2 m_i \V_i^2$ is the total energy per atom. Intuitively, the \scare{potential} term $\Jpot$ describes how the total instantaneous change in $U_i$ can be attributed to interactions with other atoms, with energy flowing between them, while the second, \scare{convective}, term $\Jconv$ describes energy being carried the individual atoms.
In the present setting, $\Jconv$ can be computed directly, as $E_i$ are available. $\Jpot$, however, presents a challenge in an \ad framework:
In principle, $\Jpot$ could be computed directly, obtaining the required partial derivatives with \ad.
% \begin{equation}
% 	\dur{i}{j}{} \hspace{1cm} i \in \Rsc, j \in \Rall \, .
% \end{equation}
However, as $\Jpot$ is neither a Jacobian-vector nor a vector-Jacobian product, this requires repeated evaluations over the input or output dimension. Even when restricting $j\in\Rsc$, which can be achieved by introducing the \mic for $\R_{ji}$ (see \SM for details), 
\begin{align}
    \Jpot^{\text{MIC}} = \sum_{i,j \in \Rsc} \left(\Rm_{ji} \left(\dur{i}{j}{} \cdot \V_j\right) \right) \,,\label{eq:J_mic}
\end{align}
computational cost of a direct implementation with \ad scales quadratically with $N$, rendering the system sizes and simulation times required for the \gk method inaccessible~\cite{lksr2023a}. We therefore consider approaches that restore linear scaling in the following.

For $\interactions{=}1$, edges can be uniquely assigned to atomic energy contributions as discussed for atomic stresses in Sec.\,\ref{sec:stress}. In this case
\begin{equation}
	\dur{}{i}{j} = \sum_{k\in\Rsc} \dur{k}{i}{j} = \dur{i}{i}{j} = \dur{i}{j}{} \,, \label{eq:ui_edge}
\end{equation}
so that
\begin{equation}
	\Jpot^{\interactions{=}1} = \sum_{ij \in \edges} \left(\R_{ji} \left(\dur{}{i}{j} \cdot \V_j\right) \right) \, , \label{eq:j_virials}
\end{equation}
which requires a single evaluation of reverse-mode \ad.

We note that the terms appearing in front of $\V_j$ also appear in the stress in \cref{eq:glp_stress_edges}. However, for a given $j$, the pre-factor \emph{cannot} be identified with the atomic stress as defined in \cref{eq:stress_atomic} -- the atomic energy being differentiated is not $U_j$, but $U_i$. The indices can only be exchanged for additive pairwise potentials; this inequivalence was recently corrected in the LAMMPS code~\cite{bbw2019t}.

This approach is not applicable for $\interactions{>}1$, since the relation in \cref{eq:ui_edge} no longer holds, and the mapping between stress contributions and heat flux contributions becomes invalid.
By using the unfolded construction, however, linear scaling can be restored regardless\cite{lksr2023a}.
Introducing auxiliary positions $\Rng_i$, which are numerically identical to the positions $\R_i$, but not used to compute $U$, and defining the energy barycenter $\Bary = \sum_{i \in \Rsc}\nolimits \Rng_i U_i$, the heat flux can be written as
\begin{align}
\Jpot^{\interactions{\geq}1} &= \sum_{j \in \Runf} \frac{\partial \Bary}{\partial \R_j} \cdot \V_j 
- \sum_{j \in \Runf}
\left(\R_j \left(\dur{}{j}{} \cdot \V_j\right) \right) \, . \label{eq:j_unf}
\end{align}
The first term requires three reverse-mode evaluations, or one forward-mode evaluation, the latter a single backward- or forward-mode evaluation. Since the overhead introduced by explicitly constructing $\Runf$ scales as $N^{2/3}$, overall linear scaling is restored, albeit with a pre-factor due to the higher number of positions to be considered.

To summarize, we have introduced two forms of the heat flux that can be implemented efficiently with \ad: \Cref{eq:j_virials}, which applies to \glps{} with $\interactions{=}1$, and \cref{eq:j_unf}, which applies for $\interactions{\geq}1$, but introduces some additional overhead. Both are equivalent to the general, quadratically-scaling, form given in \cref{eq:hf_general}, as seen in \cref{tab:hf.lj,tab:hf.snse}.

\section{Experiments}

\subsection{Lennard-Jones Argon}
\label{sec:lj.argon}

The stress formulas in \cref{eq:glp_stress_strain_direct,eq:glp_stress_strain_edges,eq:glp_stress_strain_unf,eq:glp_stress_sc_basis,eq:glp_stress_edges,eq:glp_stress_bulk} and heat flux formulas \cref{eq:J_mic,eq:j_virials,eq:j_unf} have been implemented in the \cglp package~\cite{glp} using \jax~\cite{jax}.
As a first step, we numerically verify this implementation.

To this end, the Lennard-Jones potential~\cite{l1924p} is employed, where analytical derivatives including those required for the heat flux are readily available, and implementations are included in many packages, for example the \gls{ase}~\cite{ase}.
In the \glp framework, the Lennard-Jones potential can be seen as an extreme case of a $\interactions{=}1$ \glp, where $U_i$ is composed of a sum of pair terms:
\begin{align}
    U_i &= \frac{1}{2} \sum_{j \in \nbh{i}} 4 \epsilon \left( \frac{\sigma^{12}}{r_{ij}^{12}} -\frac{\sigma^6}{r_{ij}^6} \right) \, . \label{eq:glp_lj}
\end{align}

For this experiment, parameters approximating elemental argon are used~\cite{mk2004t}.
\num{100} randomly displaced and distorted geometries, based on the \num{512}-atom $8\times8\times8$ supercell of the face-centered cubic primitive cell with lattice parameter $\SI{3.72}{\angstrom}$ and angle $\SI{60}{\degree}$ are used. Random velocities to evaluate a finite heat flux are sampled from the Boltzmann distribution corresponding to \SI{10}{K}. Additional computational details are discussed in the \SM
  
\Cref{tab:stress.lj} compares the stress formulations in \cref{eq:glp_stress_strain_direct,eq:glp_stress_strain_edges,eq:glp_stress_strain_unf,eq:glp_stress_sc_basis,eq:glp_stress_edges,eq:glp_stress_bulk} with finite differences. We report \scare{best-case} results for finite differences, choosing the stepsize that minimises the error.
In the table, the \gls{mae} and \gls{mape} with respect to the analytical ground truth are reported. All given formulations are found to be equivalent. In single precision arithmetic, the \ad-based implementations slightly outperform finite differences, in double precision, errors are similar. 

For the heat flux, finite difference approaches are not feasible. Therefore, only \ad-based implementations are shown in \cref{tab:hf.lj}. In the case of the Lennard-Jones potential, where $\interactions{=}1$, \cref{eq:J_mic,eq:j_virials,eq:j_unf} are found to be identical.

\begin{table}
\begin{tabular}{l | r r | r r}
\toprule
\multicolumn{1}{c}{}&\multicolumn{2}{c}{\textbf{Single}}&\multicolumn{2}{c}{\textbf{Double}}\\
            Equation  &  \acs{mae} (\si{eV})  &  \acs{mape} (\si{\percent})  &  \acs{mae} (\si{eV})  &  \acs{mape} (\si{\percent}) \\ 
\midrule
          Fin. diff.  &        \num{7.70e-04}  &        \num{1.04e-01}  &        \num{1.13e-06}  &        \num{1.18e-04} \\ 
\ref{eq:glp_stress_strain_direct}  &        \num{1.19e-05}  &        \num{1.79e-03}  &        \num{3.15e-06}  &        \num{3.69e-04} \\ 
\ref{eq:glp_stress_strain_edges}  &        \num{8.25e-06}  &        \num{1.27e-03}  &        \num{3.15e-06}  &        \num{3.69e-04} \\ 
\ref{eq:glp_stress_strain_unf}  &        \num{9.17e-06}  &        \num{1.36e-03}  &        \num{3.15e-06}  &        \num{3.69e-04} \\ 
\ref{eq:glp_stress_sc_basis}  &        \num{1.18e-05}  &        \num{1.79e-03}  &        \num{3.15e-06}  &        \num{3.69e-04} \\ 
\ref{eq:glp_stress_edges}  &        \num{8.22e-06}  &        \num{1.27e-03}  &        \num{3.15e-06}  &        \num{3.69e-04} \\ 
\ref{eq:glp_stress_bulk}  &        \num{9.16e-06}  &        \num{1.37e-03}  &        \num{3.15e-06}  &        \num{3.69e-04} \\ 
\bottomrule
\end{tabular}

\caption{
  \label{tab:stress.lj}
  Error in stress for Lennard-Jones argon, comparing different formulations, as well as finite differences, to analytical derivatives.
  Results are shown for both single and double precision arithmetic, and for $\stress \cdot V$ in place of $\stress$.
}
\end{table}

\begin{table}
\begin{tabular}{l | r r | r r}
\toprule
\multicolumn{1}{c}{}&\multicolumn{2}{c}{\textbf{Single}}&\multicolumn{2}{c}{\textbf{Double}}\\
                 Eq.  &  \acs{mae} (\si{eV \angstrom \per fs})  &  \acs{mape} (\si{\percent})  &  \acs{mae} (\si{eV \angstrom \per fs})  &  \acs{mape} (\si{\percent}) \\ 
\midrule
      \ref{eq:J_mic}  &        \num{2.81e-09}  &        \num{1.71e-02}  &        \num{1.47e-10}  &        \num{6.81e-04} \\ 
  \ref{eq:j_virials}  &        \num{2.84e-09}  &        \num{1.67e-02}  &        \num{1.47e-10}  &        \num{6.81e-04} \\ 
      \ref{eq:j_unf}  &        \num{2.44e-09}  &        \num{1.54e-02}  &        \num{1.47e-10}  &        \num{6.81e-04} \\ 
\bottomrule
\end{tabular}

\caption{
  \label{tab:hf.lj}
  Error in heat flux for Lennard-Jones argon, comparing different formulations to analytical derivatives.
  Results are shown for both single and double precision arithmetic.
}
\end{table}

\section{Tin Selenide with So3krates}

To investigate stress and heat flux in a practical setting, we now study tin selenide (\ch{SnSe}) using the state-of-the-art \sok \glp~\cite{fum2022q}. In contrast to other equivariant \mlps, for instantce \nequip~\cite{bmsk2022q}, \sok replaces shared equivariant feature representations by separated branches for invariant and equivariant information, whose information exchange is handled using an equivariant self-attention mechanism. By doing so, one can achieve data efficiency and extrapolation quality competitive to state-of-the-art \glps at reduced time and memory complexity.
As non-local interactions are not modeled in the \glp framework introduced in this work, global interactions are disabled in the \sok models used at present.

For these experiments, \sok models with $\interactions{=}1,2,3$ were trained on approximately \num{3000} reference calculations, comprising a number of thermalization trajectories at different volumes at \SI{300}{K}. These calculations were perfomed as part of a large-scale \gls{aigk} benchmark study by Knoop~\etal~\cite{kpsc2022t,nomad-data}. Additional details on the \mlp training can be found in the \SM

\subsubsection{Implementation of stress and heat flux}

\begin{table}
\begin{tabular}{l | r r | r r}
\toprule
\multicolumn{1}{c}{}&\multicolumn{2}{c}{\textbf{Single}}&\multicolumn{2}{c}{\textbf{Double}}\\
            Equation  &  \acs{mae} (\si{eV})  &  \acs{mape} (\si{\percent})  &  \acs{mae} (\si{eV})  &  \acs{mape} (\si{\percent}) \\ 
\midrule
\ref{eq:glp_stress_strain_direct}  &        \num{1.58e-02}  &        \num{4.40e-02}  &        \num{1.45e-04}  &        \num{2.32e-04} \\ 
\ref{eq:glp_stress_strain_edges}  &        \num{1.58e-02}  &        \num{4.38e-02}  &        \num{1.45e-04}  &        \num{2.32e-04} \\ 
\ref{eq:glp_stress_strain_unf}  &        \num{1.57e-02}  &        \num{4.33e-02}  &        \num{1.45e-04}  &        \num{2.32e-04} \\ 
\ref{eq:glp_stress_sc_basis}  &        \num{1.58e-02}  &        \num{4.40e-02}  &        \num{1.45e-04}  &        \num{2.32e-04} \\ 
\ref{eq:glp_stress_edges}  &        \num{1.58e-02}  &        \num{4.38e-02}  &        \num{1.45e-04}  &        \num{2.32e-04} \\ 
\ref{eq:glp_stress_bulk}  &        \num{1.57e-02}  &        \num{4.33e-02}  &        \num{1.45e-04}  &        \num{2.32e-04} \\ 
\bottomrule
\end{tabular}

\caption{
  \label{tab:stress.snse}
  Error in stress for tin selenide, comparing different formulations to finite differences, for \sok with $\interactions{=}2$.
  Results are shown for both single and double precision arithmetic, and for $\stress \cdot V$ in place of $\stress$. Results for other $\interactions$, which are similar to the one shown here, can be found in the \SM
}
\end{table}

\begin{table}
\begin{tabular}{l l | r r | r r}
\toprule
\multicolumn{2}{c}{}&\multicolumn{2}{c}{\textbf{Single}}&\multicolumn{2}{c}{\textbf{Double}}\\
                 Eq.  &       $\interactions$  &  \acs{mae} (\si{eV \angstrom \per fs})  &  \acs{mape} (\si{\percent})  &  \acs{mae} (\si{eV \angstrom \per fs})  &  \acs{mape} (\si{\percent}) \\ 
\midrule
  \ref{eq:j_virials}  &                   $1$  &        \num{5.78e-09}  &        \num{9.74e-04}  &        \num{1.09e-17}  &        \num{1.73e-12} \\ 
      \ref{eq:j_unf}  &                   $1$  &        \num{8.75e-08}  &        \num{2.65e-02}  &        \num{1.69e-16}  &        \num{4.31e-11} \\ 
\midrule
  \ref{eq:j_virials}  &                   $2$  &        \num{3.73e-03}  &        \num{5.84e+02}  &        \num{3.73e-03}  &        \num{5.84e+02} \\ 
      \ref{eq:j_unf}  &                   $2$  &        \num{9.36e-08}  &        \num{1.00e-02}  &        \num{1.54e-16}  &        \num{1.60e-11} \\ 
\midrule
  \ref{eq:j_virials}  &                   $3$  &        \num{1.01e-03}  &        \num{3.27e+02}  &        \num{1.01e-03}  &        \num{3.25e+02} \\ 
      \ref{eq:j_unf}  &                   $3$  &        \num{9.74e-08}  &        \num{3.04e-02}  &        \num{1.65e-16}  &        \num{2.91e-11} \\ 
\bottomrule
\end{tabular}

\caption{
  \label{tab:hf.snse}
  Error in heat flux for tin selenide, comparing different formulations to the baseline implementation in \cref{eq:J_mic} for \sok models with differing numbers of interaction steps $\interactions$.
  Results are shown for both single and double precision arithmetic.
}
\end{table}

While no analytical derivatives are available for \sok, the implementation of the stress can be verified with finite differences, and the heat flux can be checked for consistency between different implementations. Similar to Sec.\,\ref{sec:lj.argon}, we use \num{100} randomly displaced and distorted $4\times8\times8$ supercells of the \SI{0}{K} primitive cell of \ch{SnSe} for this experiment, sampling velocities from the Boltzmann distribution at \SI{10}{K} to evaluate a finite heat flux.

\Cref{tab:stress.snse} compares the stress implementations in  \cref{eq:glp_stress_strain_direct,eq:glp_stress_strain_edges,eq:glp_stress_strain_unf,eq:glp_stress_sc_basis,eq:glp_stress_edges,eq:glp_stress_bulk} with finite differences, confirming the equivalence of all implemented formulations.

\Cref{tab:hf.snse} compares the heat flux formulations in \cref{eq:j_virials} and \cref{eq:j_unf} with the baseline in \cref{eq:J_mic}, implementing the quadratically-scaling \scare{Hardy} heat flux with the \mic. For $\interactions{=}1$, all formulations are precisely equivalent. For $\interactions{>}1$, the semi-local case, only \cref{eq:j_unf} is equivalent to the \scare{Hardy} heat flux; \cref{eq:j_virials} does not apply and consequently is not equivalent, displaying large deviations.

\subsubsection{Equation of state and pressure}

\begin{figure}
  \centering
  \includegraphics[width=\columnwidth]{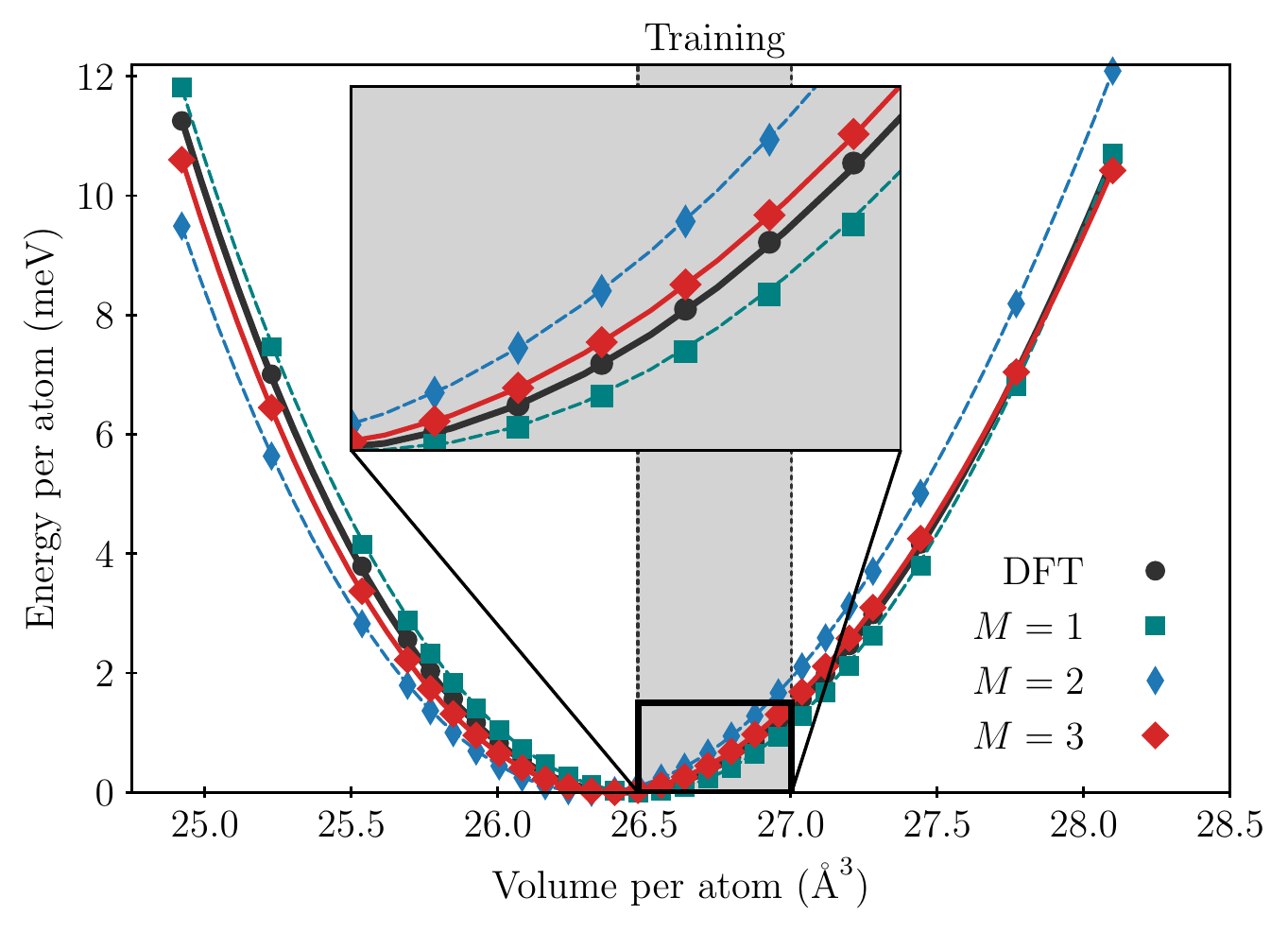}
  \caption{
    \label{fig:eos.snse.01}
    \eos (energy vs. volume) computed with PBEsol-\dft (black dots) compared to \sok with different numbers of interaction steps $\interactions$.
    The connecting lines have been obtained by fitting the Vinet \eos.
    Inset: Zoom into the region of volumes that were covered during the training indicated by the gray shading.
  }
\end{figure}

To assess the capability of \sok to predict stress- and pressure-related materials properties, we calculate energy-volume and pressure-volume curves for \ch{SnSe} to obtain an \eos of the Vinet form~\cite{vfrs1987p,hz2004p}.
The experiment is performed for unit cells that were homogeneously strained up to \SI{\pm2}{\percent} starting from the fully relaxed geometry, and relaxing the internal degrees of freedom afterwards. The energy vs. volume curves for \sok with $\interactions{=}1,2,3$ interaction steps are shown in \cref{fig:eos.snse.01} in comparison to the \dft reference using the PBEsol exchange-correlation functional with \scare{light} default basis sets in FHI-aims~\cite{FHI-aims,przb2008t}.
\sok with three interaction steps ($\interactions{=}3$) yields the best visual agreement for the energy-volume curve in \cref{fig:eos.snse.01}, and the best equilibrium volume.
To quantify the agreement, we evaluate the Vinet \eos and extract the cohesive properties equilibrium volume $V_0$,  isothermal Bulk modulus $B_0$, and its pressure derivative $B_0'$ (functional forms are given in the \SM). Results are listed in \cref{tab:eos.snse.01}. 
\begin{table}
\begin{tabular}{lrrrr}
\toprule
               & \dft                  &  $\interactions{=}1$        &  $\interactions{=}2$        &  $\interactions{=}3$        \\
$B_0$ (\si{eV/\angstrom^3}) & \num{0.230}               & \num{0.236}                & \maxf{\num{0.229} }      & \num{0.223}                \\
$B_p$ (\si{eV/\angstrom^6}) & \num{5.576}                & \num{2.830}                & \maxf{\num{4.815} }      & \num{7.118}                \\
$V_0$ (\si{\angstrom^3})    & \num{26.429}               & \num{26.489}               & \num{26.322}               & \maxf{\num{26.388}}      \\ \midrule
%               & \multicolumn{1}{l}{} & \multicolumn{1}{l}{} & \multicolumn{1}{l}{} & \multicolumn{1}{l}{} \\
Error $B_0$ (\si{\percent})  & --\hspace{0.2cm}                    & \num{2.58}               & \maxf{\num{-0.47} }    & \num{-3.14}              \\
Error $B_p$ (\si{\percent})  & --\hspace{0.2cm}                    & \num{-49.25}             & \maxf{\num{-13.64}}    & \num{27.66}              \\
Error $V_0$ (\si{\percent})  & --\hspace{0.2cm}                    & \num{0.22}               & \num{-0.40}              & \maxf{\num{-0.16} }    \\
\bottomrule
\end{tabular}
\caption{
  \label{tab:eos.snse.01}
  Cohesive properties of \ch{SnSe} for PBEsol-\dft, and \sok with different numbers of interaction steps $\interactions$ obtained via the Vinet \eos. Best values are highlighted. Values are in good agreement with existing literature~\cite{lccp2018p}.
}
\end{table}
For \sok with three interaction steps ($\interactions{=}3$), the predicted volume deviates by \SI{-0.16}{\percent} from the \dft reference, and the deviation of the bulk modulus is \SI{-3.14}{\percent}. 
A larger error is seen for the pressure derivative of the bulk modulus, $B_0'$, which deviates by \SI{27.7}{\percent}, indicating worse agreement further away from the training region.
Overall, the agreement between \dft and \sok when predicting cohesive properties can be considered satisfactory. The energy-volume predictions are very good, and transfer even to volumes that are larger or smaller then the ones seen during training.

\begin{figure}
  \centering
  \includegraphics[width=\columnwidth]{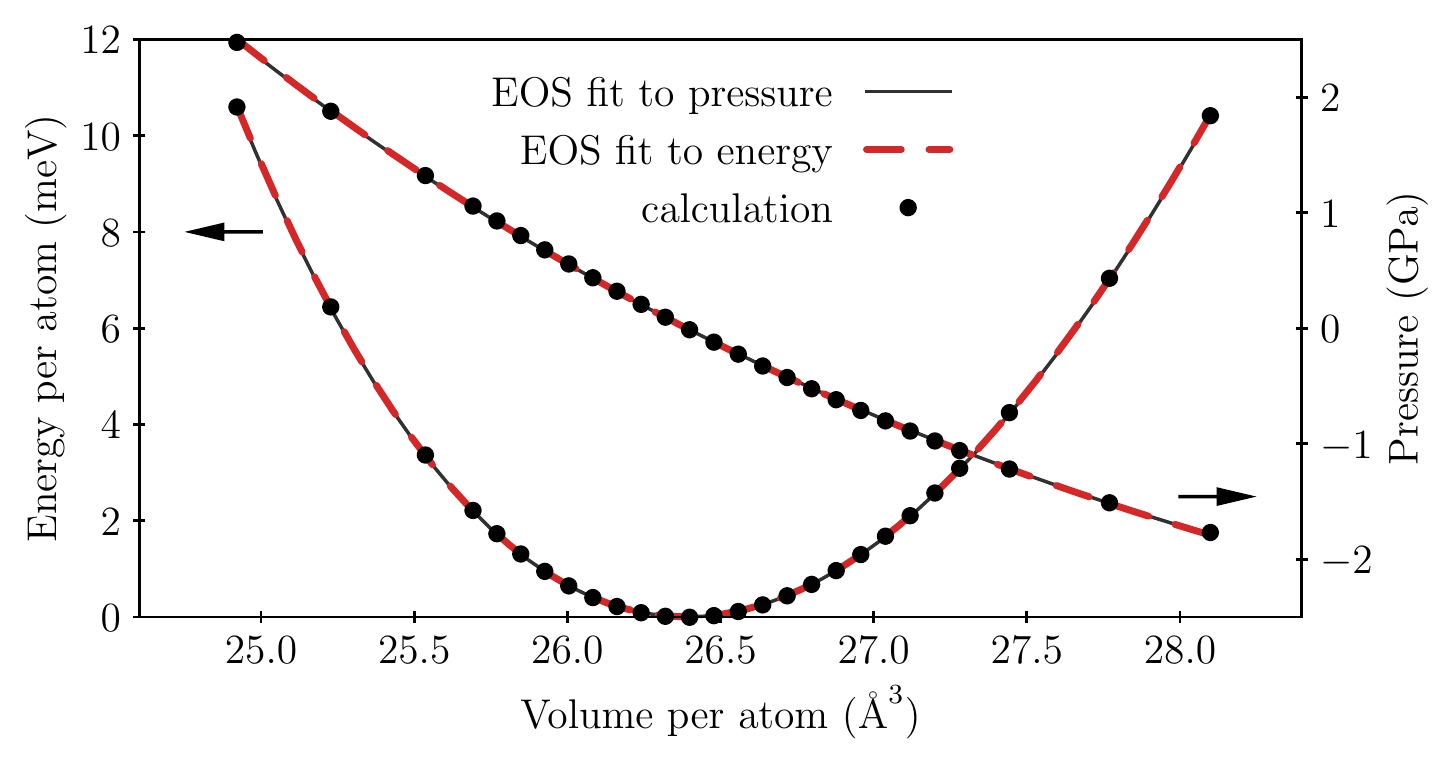}
  \caption{
    \label{fig:eos.snse.02}
    Comparison of \eos obtained by fitting energy-volume (red dashed) and pressure-volume (black solid) data. The results are in perfect agreement.
  }
\end{figure}

Finally we check the internal consistency of energy and stress predictions with \sok by fitting energy-volume and pressure-volume curves and verify that they yield identical parameters for the \eos. Results are shown in \cref{fig:eos.snse.02}. As seen there, results are in perfect agreement, which verifies that stress and resulting pressure are consistent with the underlying energy function as expressed in \cref{eq:stress}.

\begin{table*}
\begin{tabular}{llr|rrr}
\toprule
              Source  &                Method  &       $\kappa$ (W/mK)  &   $\kappa_{\text{x}}$  &   $\kappa_{\text{y}}$  &   $\kappa_{\text{z}}$ \\ 
\midrule
           This work  &  \sok, $\interactions{=}1$  &   \num{0.99 \pm 0.10}  &   \num{0.53 \pm 0.03}  &   \num{1.31 \pm 0.13}  &   \num{1.12 \pm 0.12} \\ 
\hspace{0.5cm}\textquotedbl  &  \sok, $\interactions{=}2$  &   \num{1.13 \pm 0.07}  &   \num{0.48 \pm 0.04}  &   \num{1.59 \pm 0.07}  &   \num{1.20 \pm 0.07} \\ 
\hspace{0.5cm}\textquotedbl  &  \sok, $\interactions{=}3$  &   \num{1.13 \pm 0.10}  &   \num{0.56 \pm 0.05}  &   \num{1.56 \pm 0.15}  &   \num{1.32 \pm 0.16} \\ 
Brorsson~\etal~\cite{bhke2021t}  &             \acs{fcp}  &            \num{1.12}  &            \num{0.57}  &            \num{1.46}  &            \num{1.32} \\ 
Liu~\etal~\cite{lqzg2021q}  &             \acs{mlp}  &   \num{0.86 \pm 0.13}  &   \num{0.57 \pm 0.05}  &   \num{1.25 \pm 0.24}  &   \num{0.76 \pm 0.08} \\ 
Knoop~\etal~\cite{kpsc2022t}  &  \acs{dft} (extrapolated)  &   \num{1.40 \pm 0.39}  &                    --  &                    --  &                    -- \\ 
Review by Wei~\etal~\cite{wbcr2016t}  &           Experiments  &  \num{0.45} to \num{1.9}  &                    --  &                    --  &                    -- \\ 
\bottomrule
\end{tabular}

\caption{
  \label{tab:snse_kappa}
  Thermal conductivity $\kappa$ of tin selenide at \SI{300}{K}.
}
\end{table*}

\subsubsection{Thermal Conductivity}

Finally, we proceed to \gk calculations, following the approach outlined in our previous work\cite{kcs2022t,lksr2023a}.
Finding that simulation cells with $N{=}\num{864}$ atoms and a simulation duration of $\td{=}\SI{1}{ns}$ yield converged results (see \SM), we run \num{11} \md simulations using the \texttt{glp} package and \sok, computing $\J$ at every step.

\Cref{tab:snse_kappa} shows the result: For $\interactions{=}2,3$, \sok in excellent agreement with the work by Brorsson~\etal~\cite{bhke2021t}, which uses a \gls{fcp}.
A larger difference is observed with the work by Liu~\etal~\cite{lqzg2021q}, who, however, use the PBE exchange-correlation functional, as opposed to PBEsol, which was used for the present work.
The observed thermal conductivity is consistent with experiments, as well as the size-extrapolated \dft result of Knoop~\etal~\cite{kpsc2022t}.
The anisotropy of $\tk$ in \ch{SnSe} is captured as well.
Overall, \sok with more than one interaction step is able to capture the converged thermal conductivity of \ch{SnSe}, using only training data from the thermalization step of a full \gls{aigk} workflow; long-running equilibrium \emph{ab initio} MD simulations, the bottleneck of the \gk method, have been avoided.

\section{Conclusion}

We demonstrated that the stress and heat flux can be computed efficiently with \ad for potentials based on a graph of atom-pair vectors, which we termed \glps, and provided example implementations in the \cglp package~\cite{glp}.
Numerical experiments for Lennard-Jones argon and tin selenide with the \sok \glp, verified that these quantities are computed correctly and consistently.
The equivariant \sok \glp was shown to predict cohesive properties and thermal conductivity of \ch{SnSe} in good agreement with \dft, other \mlps, and experiments, confirming the practical relevance of computational access to stress and heat flux.

This work enables the use of a large class of recently developed \mlps, those which can be described in the \glp framework, in computational materials science, and in particular for the calculation of thermal conductivities using the \gk method.
For \glps implemented with \jax, the \cglp package is provided to enable the calculation of stress and heat flux without requiring further implementation efforts.

\vspace{1cm}

\section*{Data and Code Availability}

The data and code that support the findings of this study are available at \href{https://doi.org/10.5281/zenodo.7852530}{doi:10.5281/zenodo.7852530} and at \href{https://github.com/sirmarcel/glp-archive}{https://github.com/sirmarcel/glp-archive}. The \cglp package is available at \href{https://github.com/sirmarcel/glp}{https://github.com/sirmarcel/glp}. The \sok model is implemented in \cmlff, available at \href{https://github.com/thorben-frank/mlff}{https://github.com/thorben-frank/mlff}.
 % \gk calculations were run with \cgkx, available at \href{https://github.com/sirmarcel/gkx}{https://github.com/sirmarcel/gkx}.
Further information can be found in the \SM and at \href{https://marcel.science/glp}{https://marcel.science/glp}.

\begin{acknowledgments}
M.F.L. was supported by the German Ministry for Education and Research BIFOLD program (refs. 01IS18025A and 01IS18037A), and by the TEC1p Project (ERC Horizon 2020 No. 740233). 
M.F.L. would like to thank Samuel Schoenholz and Niklas Schmitz for constructive discussions, Shuo Zhao for feedback on the manuscript, and acknowledges contributions by Fabian Nagel and Adam Norris.
J.T.F acknowledges support from the Federal Ministry of Education and Research (BMBF) and BIFOLD program (refs. 01IS18025A and 01IS18037A).
F.K. acknowledges support from the Swedish Research Council (VR) program 2020-04630, and the Swedish e-Science Research Centre (SeRC).
The computations were partially enabled by the Berzelius resource provided by the Knut and Alice Wallenberg Foundation at the National Supercomputer Centre (NSC) and by resources provided by the National Academic Infrastructure for Supercomputing in Sweden (NAISS) at NSC partially funded by the Swedish Research Council through grant agreement no. 2022-06725.
Xuan Gu at NSC is acknowledged for technical assistance on the Berzelius resource.
\end{acknowledgments}

\section*{Author Declarations}

\subsection*{Conflict of Interest}

The authors have no conflicts to disclose.

\subsection*{Author Contributions}
% https://groups.niso.org/higherlogic/ws/public/download/26466/ANSI-NISO-Z39.104-2022.pdf

\noindent
\textbf{Marcel F. Langer}: Conceptualization; Writing -- original draft (lead); Writing -- review \& editing (lead); Software (lead); Visualization (equal); Investigation (equal); Project administration.
\textbf{J. Thorben Frank}: Writing -- original draft (supporting); Writing -- review \& editing (supporting); Software (supporting); Investigation (equal).
\textbf{Florian Knoop}: Writing -- original draft (supporting); Writing -- review \& editing (supporting); Visualization (equal); Investigation (equal).

\bibliography{babel_short,custom}

\end{document}

\preprint{AIP/123-QED}

\title{Supplementary Material:\\Stress and heat flux via automatic differentiation}

\author{Marcel F. Langer}
\email{mail@marcel.science}
\affiliation{Machine Learning Group, Technische Universit{\"a}t Berlin, 10587 Berlin, Germany}
\affiliation{BIFOLD -- Berlin Institute for the Foundations of Learning and Data, Berlin, Germany}
\affiliation{The NOMAD Laboratory at the Fritz Haber Institute of the Max Planck Society and Humboldt University, Berlin, Germany}

\author{J. Thorben Frank}
\affiliation{Machine Learning Group, Technische Universit{\"a}t Berlin, 10587 Berlin, Germany}
\affiliation{BIFOLD -- Berlin Institute for the Foundations of Learning and Data, Berlin, Germany}

\author{Florian Knoop}
\affiliation{Theoretical Physics Division, Department of Physics, Chemistry and Biology (IFM), Linköping University, SE-581 83 Linköping, Sweden}

\date{\today}%

\maketitle

\section{Heat Flux with Minimum Image Convention}

We assume that $\effcutoff$ is chosen such that each atom $i$ can interact with at most one replica of each other atom $j$, i.e., $\effcutoff$ does not exceed half the smallest distance between opposing faces of the simulation cell.
In that case, the partial derivative in Eq.~17 of the manuscript can be computed with respect to the equivalent position in the simulation cell.
However, the atom-pair vector $\R_{ji}$ must still connect $\R_i$ in the simulation cell with the respective position $\R_j$, which may be a replica.
This can be ensured by adopting the \mic, yielding
\begin{equation}
  \Jpot^{\text{MIC}} \defas \sum_{i,j \in \Rsc} \left(\Rm_{ji} \left(\dur{i}{j}{} \cdot \V_j\right) \right)
\end{equation}
While this form of the heat flux requires no replica positions, it is also unsuitable for the efficient implementation with \ad: As different factors are multiplied with each entry of the Jacobian $\indur{i}{j}{}$, the evaluation of this heat flux requires the computation of the explicit full Jacobian, leading to quadratic scaling.

However, as it requires no modification in the implementation of a given potential, and can be implemented directly, albeit inefficiently, with \ad, we use $\Jpot^{\text{MIC}}$ as baseline for the development of more specialized approaches.

\section{Lennard-Jones Argon}

\subsection{Implementation}

The Lennard-Jones potential used in this work is based on the \scare{smooth} implementation in \ase~\cite{ase}, where the standard pair potential is multiplied with a cutoff function
\begin{equation}
  \fcutoff(r) = \begin{cases}
    1 &  \text{for } r < \onset \\
    \frac{(\cutoff^2 - r^2)^2 (\cutoff^2 + 2 r^2 - 3 \onset^2)}{(\cutoff^2 - \onset^2)^3} & \text{for } \onset \leq r \leq \cutoff \\
    0 &  \text{for } r > \cutoff \, ,
  \end{cases}
\end{equation}
ensuring that energies and forces decay continuously to zero as atoms approach $\cutoff$.

This was implemented in \cglp, and energy predictions were verified to ensure that \cglp and \ase yield identical results.

\subsection{Test of stress and heat flux}
\label{ssec:lj.stress.hf}

In order to verify the approach described in this work, and to test the \cglp framework, stress and heat flux were compared with the implementation in \ase, where derivatives are computed analytically.
The experiment consists of computing these properties for \num{100} randomly perturbed simulation cells of Lennard-Jones argon, starting from an $8\times8\times8$ supercell of the face-centred cubic primitive cell with lattice parameter $\SI{3.72}{\angstrom}$ and angle $\SI{60}{\degree}$.

Positions are then modified with perturbations drawn from a normal distribution with $\sigma{=}\SI{0.01}{\angstrom}$; a random strain with each component drawn from a uniform distribution over $[\SI{-1}{\percent}, \SI{1}{\percent}]$ is also applied. Velocities, required for the heat flux, are drawn from a Boltzmann distribution corresponding to \SI{10}{K}. For the heat flux, only $\Jpot$ is computed, as $\Jconv$ is identical for all implementations.

\section{Tin Selenide}

\subsection{Implementation of test and heat flux}

For this experiment, the relaxed \SI{0}{K} primitive cell of \ch{SnSe}, obtained from the work by Knoop~\etal~\cite{kpsc2022t}, was extended to a $4\times 8\times 8$ supercell, ensuring that $\effcutoff$ does not exceed half length of the smallest lattice vector.
Otherwise, the experiment proceeded identically to the one described in \cref{ssec:lj.stress.hf}.

\Cref{tab:stress.snse.m1,tab:stress.snse.m3} contain comparisons of the \cglp stress with finite differences for $\interactions{=}1$ and $\interactions{=}3$. Results are similar to the $\interactions{=}2$ case.

\begin{table}
\begin{tabular}{l | r r | r r}
\toprule
\multicolumn{1}{c}{}&\multicolumn{2}{c}{\textbf{Single}}&\multicolumn{2}{c}{\textbf{Double}}\\
            Equation  &  \acs{mae} (\si{eV})  &  \acs{mape} (\si{\percent})  &  \acs{mae} (\si{eV})  &  \acs{mape} (\si{\percent}) \\ 
\midrule
10  &        \num{1.42e-02}  &        \num{4.33e-02}  &        \num{1.24e-04}  &        \num{2.29e-04} \\ 
11  &        \num{1.42e-02}  &        \num{4.32e-02}  &        \num{1.24e-04}  &        \num{2.29e-04} \\ 
12  &        \num{1.42e-02}  &        \num{4.29e-02}  &        \num{1.24e-04}  &        \num{2.29e-04} \\ 
13  &        \num{1.42e-02}  &        \num{4.33e-02}  &        \num{1.24e-04}  &        \num{2.29e-04} \\ 
14  &        \num{1.42e-02}  &        \num{4.32e-02}  &        \num{1.24e-04}  &        \num{2.29e-04} \\ 
15  &        \num{1.42e-02}  &        \num{4.29e-02}  &        \num{1.24e-04}  &        \num{2.29e-04} \\ 
\bottomrule
\end{tabular}

\caption{
  \label{tab:stress.snse.m1}
  Error in stress for tin selenide, comparing different formulations to finite differences, for \sok with $\interactions{=}1$.
  Results are shown for both single and double precision arithmetic, and for $\stress \cdot V$ in place of $\stress$.
}
\end{table}

\begin{table}
\begin{tabular}{l | r r | r r}
\toprule
\multicolumn{1}{c}{}&\multicolumn{2}{c}{\textbf{Single}}&\multicolumn{2}{c}{\textbf{Double}}\\
            Equation  &  \acs{mae} (\si{eV})  &  \acs{mape} (\si{\percent})  &  \acs{mae} (\si{eV})  &  \acs{mape} (\si{\percent}) \\ 
\midrule
10  &        \num{2.07e-02}  &        \num{4.86e-02}  &        \num{1.91e-04}  &        \num{3.71e-04} \\ 
11  &        \num{2.07e-02}  &        \num{4.87e-02}  &        \num{1.91e-04}  &        \num{3.71e-04} \\ 
12  &        \num{2.06e-02}  &        \num{4.86e-02}  &        \num{1.91e-04}  &        \num{3.71e-04} \\ 
13  &        \num{2.07e-02}  &        \num{4.86e-02}  &        \num{1.91e-04}  &        \num{3.71e-04} \\ 
14  &        \num{2.07e-02}  &        \num{4.87e-02}  &        \num{1.91e-04}  &        \num{3.71e-04} \\ 
15  &        \num{2.06e-02}  &        \num{4.86e-02}  &        \num{1.91e-04}  &        \num{3.71e-04} \\ 
\bottomrule
\end{tabular}

\caption{
  \label{tab:stress.snse.m3}
  Error in stress for tin selenide, comparing different formulations to finite differences, for \sok with $\interactions{=}3$.
  Results are shown for both single and double precision arithmetic, and for $\stress \cdot V$ in place of $\stress$.
}
\end{table}

\subsection{Evaluating the potential}

This section presents additional results with \sok and \ch{SnSe}, which were used to assess the quality of the \mlps used in the work.

\begin{figure}
  \centering
  \includegraphics[width=\columnwidth]{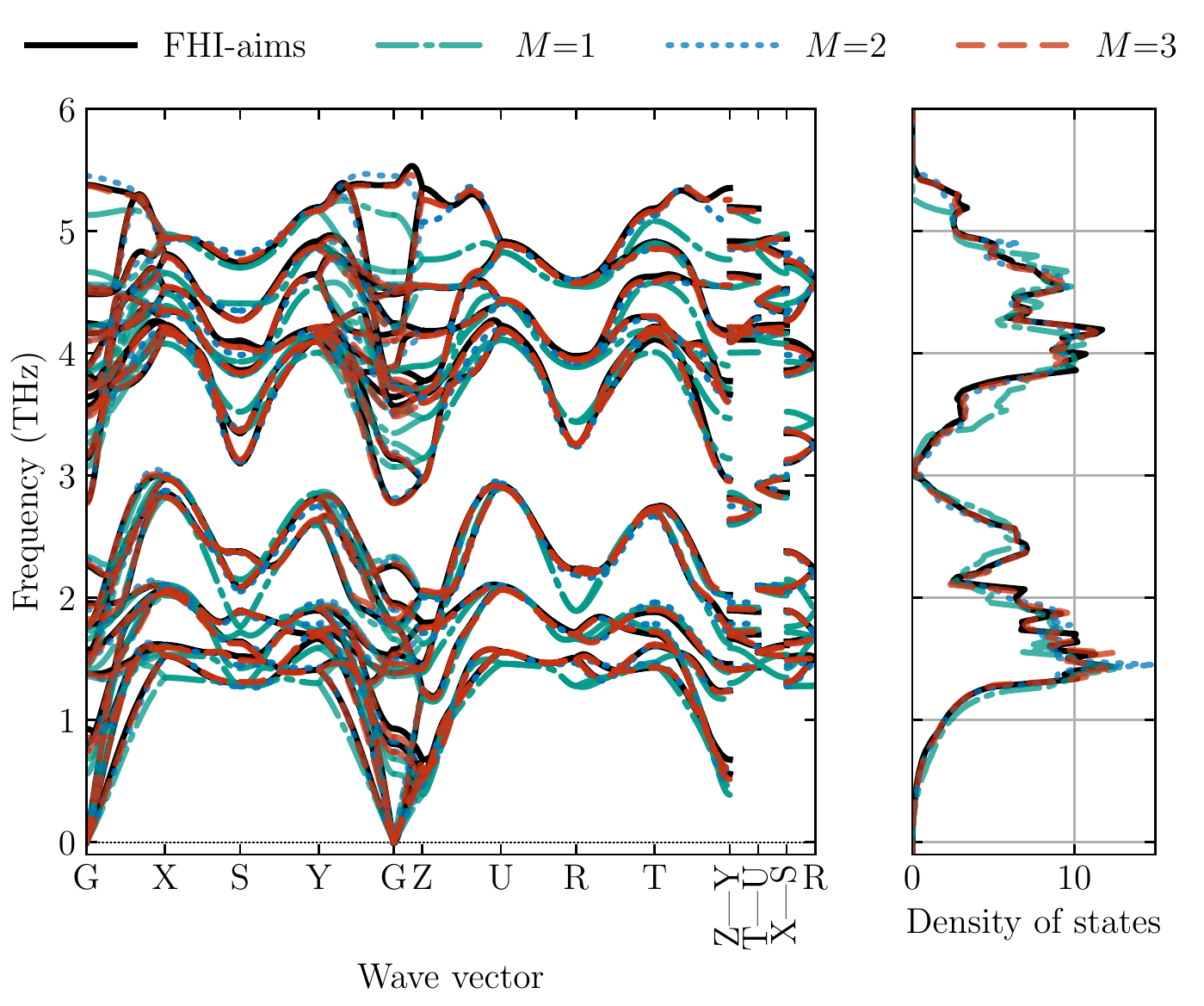}
  \caption{
    \label{fig:si.snse.phonons}
    Phonon band structure and density of states for tin selenide for \sok models with $\interactions{=}1,2,3$ compared with FHI-aims. Results are reported for a supercell with \num{256} atoms.
  }
\end{figure}

\Cref{fig:si.snse.phonons} shows the phonon band structure for \sok with different $\interactions$ compared with the \dft reference. Agreement between \dft and \sok increases systematically with $\interactions$; satisfactory agreement is obtained from $\interactions{=}2$ onwards.

\begin{figure}
  \centering
  \includegraphics[width=\columnwidth]{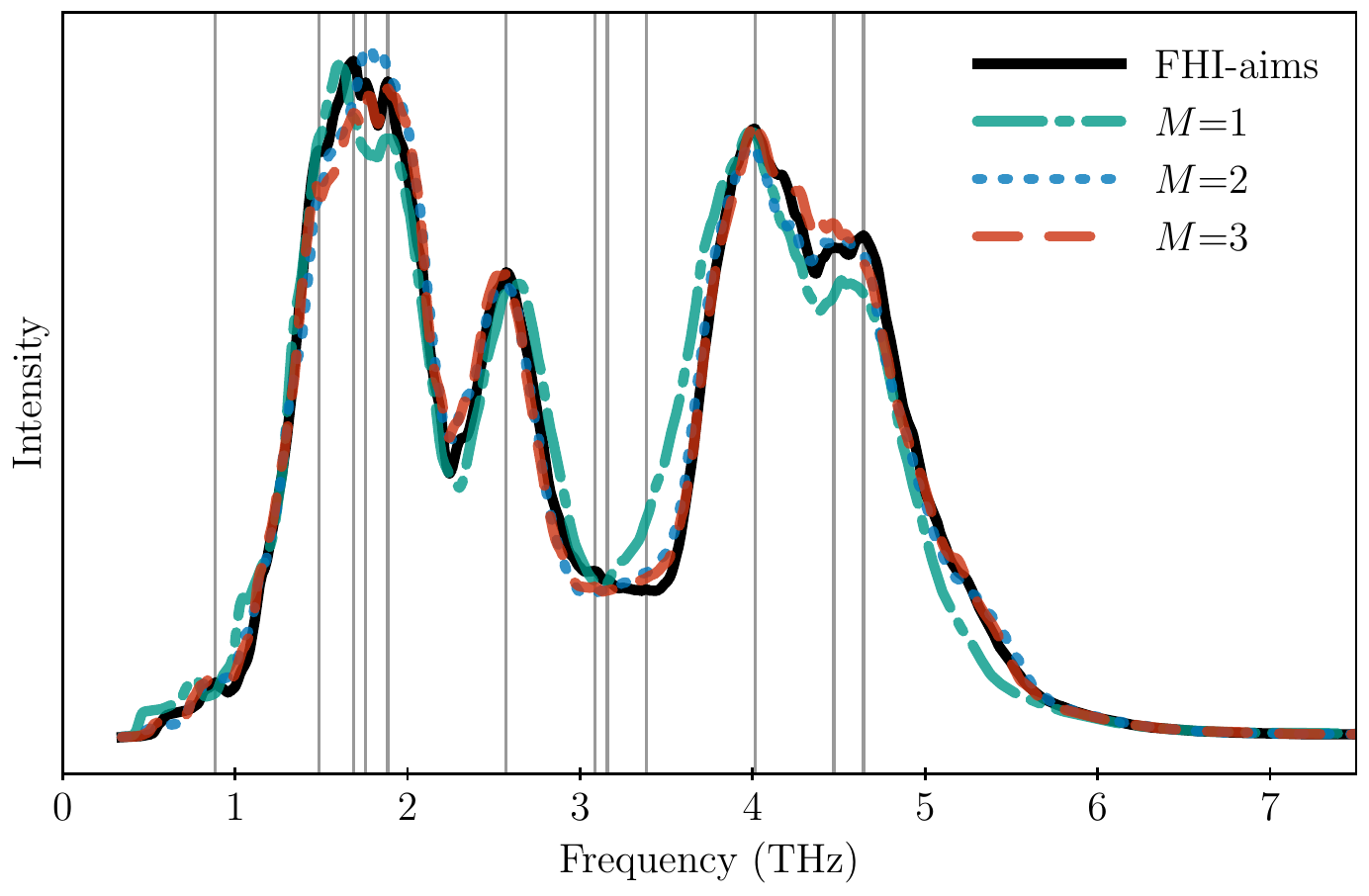}
  \caption{
    \label{fig:si.snse.vdos}
    Vibrational density of states for tin selenide at \SI{300}{K} for \sok models with $\interactions{=}1,2,3$ compared with FHI-aims. Results are reported for a trajectory with \SI{30}{ps} duration in a supercell with \num{256} atoms, started from identical initial configurations.
    Vertical lines indicate prominent peaks in the FHI-aims result.
  }
\end{figure}

The vibrational density of states can be seen in \cref{fig:si.snse.vdos}. Overall features are captured for all values of $\interactions$. Performance is slightly improved for $\interactions{=}2,3$, which behave similarly.

In line with observations in the main text, $\interactions{=}2$ can be considered sufficient to model vibrational properties. While slight improvements are obtained with $\interactions{=}3$, the impact on thermal conductivity is negligible. Energy-volume curves, on the other hand, are improved by $\interactions{=}3$.

\subsection{Green-Kubo workflow and convergence}

As the \gk method is formally valid in the thermodynamic limit, the convergence of $\kappa$ with respect to simulation size and duration must be considered. Additional considerations arise from the use for noise reduction in the resulting \gls{hfacf}.
In this work, following previous work~\cite{kcs2022t,lksr2023a} a lowpass filter with frequency \SI{1}{THz} is applied to the integrated \gls{hfacf}, which is then differentiated via finite differences to yield a smooth \gls{hfacf} from which the integration cutoff can be determined as the first zero crossing. No further noise reduction is performed; no gauge terms are removed and the full heat flux $\J = \Jpot + \Jconv$ is computed at all timesteps.

Eleven trajectories with timestep $\Delta t = \SI{4}{fs}$ are used throughout, initialised from snapshots from a \SI{0.2}{ns} $NVT$ thermalization trajectory using the Langevin thermostat implemented in \ase~\cite{ase}. The primitive cell at \SI{300}{K} from the work by Knoop~\etal~\cite{kpsc2022t} was used, creating $n\times2n\times2n$ supercells to obtain simulation cells at different $N=32n^3$.

\begin{figure}
  \centering
  \includegraphics[width=\columnwidth]{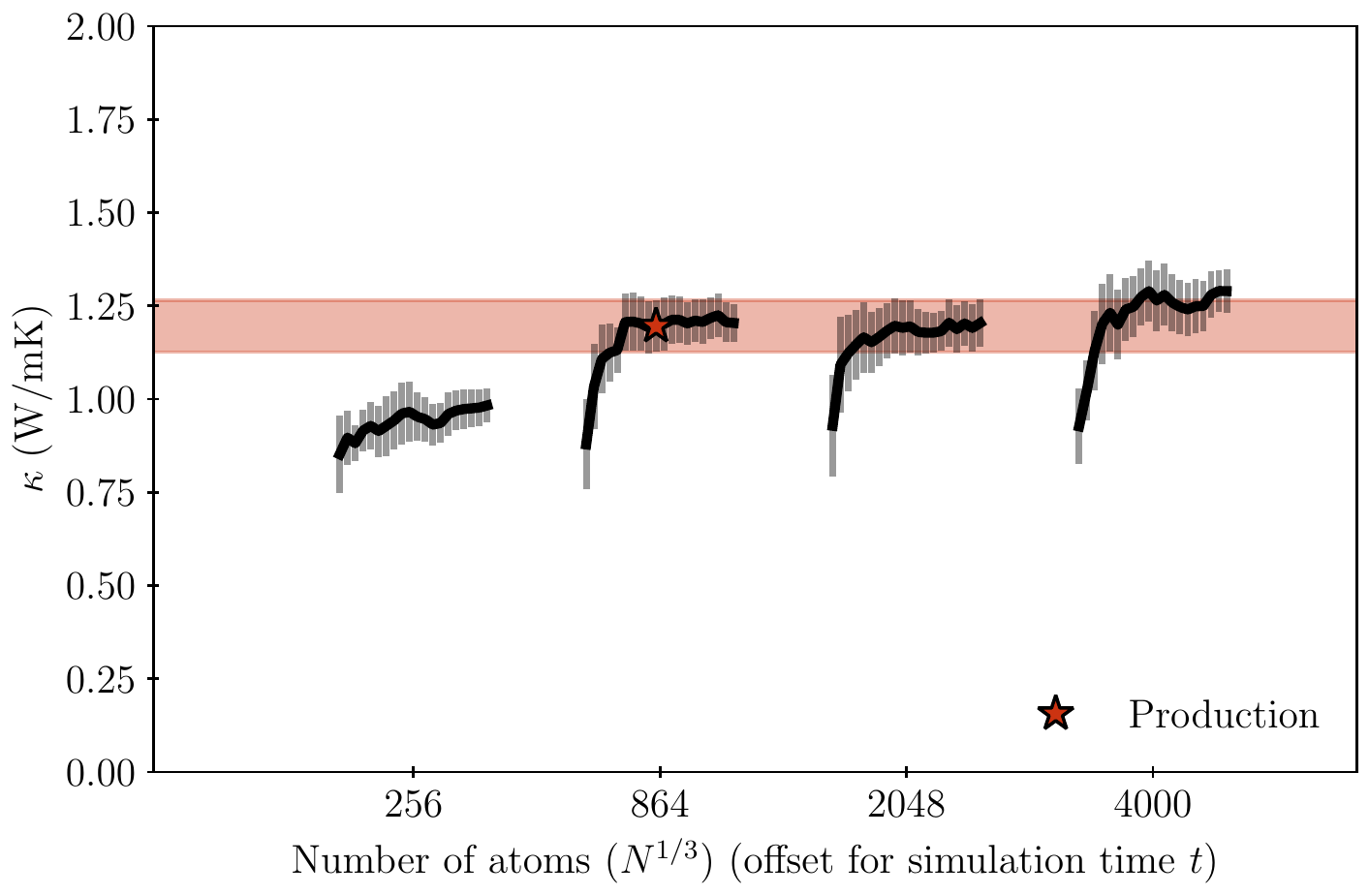}
  \caption{
    \label{fig:si.snse.convergence}
    $\kappa$ for tin selenide at \SI{300}{K} for different choices of simulation cell size $N$ and simulation duration $\td$ for \sok with $\interactions{=}2$.
    Error bars indicate the standard error across trajectories.
    $N$ is shown as $N^{1/3}$, which is proportional to the length scale of the simulation cell.
    For each choice of $N$, $\td$ from \SIrange{0.1}{4.0}{ns} are shown with a horizontal offset.
    \scare{Production} settings $\snsecc$ are indicated; the associated standard error is shown as a shaded band.
  }
\end{figure}

\Cref{fig:si.snse.convergence} displays the convergence of $\kappa$ with respect to simulation cell size $N$ and simulation duration $\td$, for \sok with $\interactions{=}2$. No significant increases in $\kappa$ are observed for simulation cell sizes above $N{=}864$ and simulation durations above $\td{=}\SI{2}{ns}$. We therefore choose $\snsecc$ as \scare{production} settings for this work.

\subsection{Model and training}
Here we use the \sok \mpnn, which is implemented in the \cmlff package~\cite{mlff}. The number of interaction steps $M$ is varied from $M=1$ up to $M=3$, whereas we fix the cutoff radius to $\cutoff=\SI{5}{\angstrom}$, the embedding dimension to $F=132$ and the maximal degree in the equivariant branch to $l_{\text{max}} = 3$. Non-local corrections are not used. After $M$ message-passing updates the final embeddings are put through a two layered \ffnn with \textsc{silu} (sigmoid linear unit) non-linearity to obtain per-atom energies. The total potential energy is given as the sum of the per-atom energies.

The model is trained by minimizing a joint loss of the potential energy and per-atom forces
with loss weightings of 0.01 and 0.99, respectively using the ADAM optimizer~\cite{kb2014m}. In total 3000 reference structures of tin selenide are used for training, of which 600 are reserved for validation. The training is stopped after 2500 epochs, with a batch size of 10. After each epoch, the performance of the current model is evaluated on the validation data and the best-performing model is saved for production. The initial learning is set to $10^{-3}$ and is reduced every $100$k steps using exponential learning rate decay with a decay factor of $0.7$. No early stopping is employed. Training times on a single NVIDIA A100 40GB GPU range from $2$h$54$min for $M=1$ up to $6$h$46$min for $M=3$.

\section{Equation of State (EOS)}

To describe cohesive properties, we use the Vinet equation of state~\cite{rsgf1984p,vfsr1986p,vsfr1987p,vfrs1987p} for pressure $p$ as function of volume $V$:
\begin{align}
  p(V)=\frac{3 B_0}{X^2}(1-X) \mathrm{e}^{\eta(1-X)}~,
  \label{eq:vinet.p}
\end{align}
with
\begin{align}
  X=\left[\frac{V}{V_0}\right]^{\frac{1}{3}} \quad \text { and } \quad \eta=\frac{3}{2}\left(B_0^{\prime}-1\right)~,
  \label{eq:vinet.variables}
\end{align}
where $V_0$ is the volume at vanishing pressure, $B_0$ the bulk modulus and $B_0'$ its volume derivative.
The respective energy function that fulfills $p = - {\rm d} E / {\rm d} V$ is given by~\cite{hz2004p}
\begin{align}
  E(X)
  &= E_0 + \frac{2 B_0 V_0}{\left(B_0^{\prime}-1\right)^2} \left(2-\left(5+3 B_0^{\prime}(X-1)-3 X\right) {\rm e}^{\eta (1-X)}\right)~.
  \label{eq:vinet.E}
\end{align}
These functions are fitted to the computed values for $E$ or $p$ and $V$ to obtain $V_0$, $B_0$, $B_0'$, and the minimal energy $E_0$ in the case of \cref{eq:vinet.E}.

\bibliography{babel_short,custom}